\documentclass[12pt]{article}
\usepackage{amssymb}
\usepackage{epsfig,caption,graphicx,eepic,epic}
\usepackage{pstricks}
\usepackage{amsmath}
\usepackage{multido}

\begin{document}
\def\be{\begin{equation}}
\def\ee{\end{equation}}
\def\ba{\begin{eqnarray}} 
\def\ea{\end{eqnarray}}
\def\nn{\nonumber}

\newcommand{\bbf}{\mathbf}
\newcommand{\rrm}{\mathrm}

\title{The Berry phase: a topological test for the spectrum structure \\
of frustrated quantum spin systems \\}

\author{
Jean Richert
\footnote{E-mail address up to December 31, 2008: richert@lpt1.u-strasbg.fr}\\ 
Laboratoire de Physique Th\'eorique, UMR 7085 CNRS/ULP,\\
Universit\'e Louis Pasteur, 67084 Strasbourg Cedex,\\ 
} 
 
\date{\today}
\maketitle 
\begin{abstract}
The nature of the low energy spectrum of frustrated quantum spin systems is investigated by means
of a topological test introduced by Y. Hatsugai~\cite{hat1} which enables to infer the possible
existence or absence of a gap between the ground state and excited states of these systems. The test 
relies on the determination of an order parameter which is a Berry phase. The structure of the spectra 
of even and odd-legged systems in $2d$ and $3d$ is analysed. Results are confronted with previous 
work.  


\end{abstract} 
\maketitle
PACS numbers: 3.65.Vf, 71.10.-w, 71.27.+a 
\vskip .2cm
Keywords: Topological properties of quantum spin systems - Berry phase - Gapped spectra.\\

{\it Introduction}.

The structure of the low energy spectra of quantum spin systems is of prime importance for the 
understanding of specific phenomena such as superconductivity at high $T_c$. Investigations on this 
subject which went on for many years have concentrated in the recent past on specific structures, in 
particular the presence of ladders and stripes in $2d$ superconducting material, see f.i.
~\cite{hot,tas,hay,tran}. It has been observed that systems which expectedly behave like 2-leg 
ladders show a gap between the ground state and the first excited state~\cite{dag,dell} whereas those 
which behave like 3-leg ladders possess a continuous spectrum~\cite{azu}. Theoretical investigations 
have been developed in order to study the properties of the ground state and low energy states with a 
particular interest for the existence or absence of such a gap. An important step was performed by 
Haldane who conjectured that the spectra of Heisenberg antiferromagnetic chains are gapless if the 
spins are half-integers and show a gap if the spins are integers~\cite{hald1,hald2}. This conjecture 
has been followed by a large amount of work extending to $2d$ and higher dimensional spin networks,
among them variational approaches ~\cite{ric1,ric2}. However the problem of low energy properties of 
these systems has not yet been completely settled.\\  

Topological concepts often work as efficient tools in the investigation of the properties of physical 
systems like those described by quantum models. Following the work of Wen ~\cite{wen1}
a considerable amount of investigations has been performed which concerns the structure of specific
systems, phenomena like topological phase transitions~\cite{wen2} and entanglement properties 
which may characterize them ~\cite{wen3,kit1}.\\

The Berry phase (BP) is a genuine topological quantum concept. It enters the phase of the wave function 
of a physical system governed by a set of parameters which vary slowly along a path $C$~\cite{berr}.
Parallel transport~\cite{goss} can be considered as its classical equivalent. It has been 
used as a universal concept in different fields of physics and works, f. i. in the description of
spinor systems, neutron interference processes, the Jahn-Teller and Aharonov-Bohm effects. Matrix
generalizations for the description of non-abelian phases have been introduced by Wilczek and Zee 
~\cite{wilc}. Recently a BP has been experimentally detected in the observation of the 
quantum Hall effect in graphenes ~\cite{zhan} and used as a revelator of the existence of entangled 
systems~\cite{zhao,ryu} and phase transitions~\cite{binz, shil}.\\ 

The geometrical essence of BP is of particular interest in connection with symmetry properties 
of quantum systems. It was introduced in this context by Hatsugai and colls.~\cite{hat2,hat3,hat4,
hat5, hat6} as a possible tool in order to characterize the properties of the low energy spectrum of 
spin systems. We use this property here in order to revisit the problem of frustrated ladder systems 
mentioned above.\\

{\it Theory}.\\

We recall the main lines of the argument of ref.~\cite{hat6} which leads to the introduction of a
constrained BP and the characterization of spectra.

The BP is finite as long as the wave function to which it belongs is nondegenerate~\cite{berr,shil}, 
degeneracy corresponds to the existence of a quantum phase transition. Consider a system which is 
not located at such a transition point and more generally does not show a degeneracy in its ground state. 
If the system shows specific symmetries (translation, reflection, rotation,...invariance) the BP of 
the considered wave function must be such that this invariance is respected when acting with the 
corresponding symmetry operator, hence it should obey a specific constraint. If this condition can be 
realized for a finite BP the system will necessarily show a gap after the symmetry operation. If it 
is not the case, one must conclude that the spectrum cannot show a gap, i. e. it must be continuous.\\

More precisely consider a general Hamiltonian

\ba
H = \sum_{(i,j)} J_{ij} \vec S_{i} \vec S_{j} 
\label{eq1}   
\ea
Apply a local twist of angle $\phi$ on the part corresponding to the sites $(k,l)$ and define

\ba
H_{kl}(\phi) = J_{kl}[1/2(exp(i\phi) S^{+}_{k} S^{-}_{l} + h.c.) + S^{z}_{k}S^{z}_{l}] +
\sum_{(i,j)\not= (kl)} J_{ij}\vec S_{i} \vec S_{j}
\label{eq2}   
\ea
with $\phi \in [0,2\pi]$.

By using a unitary transformation on the local site $k$, $U_{k}(\phi) = exp(iS-S^{z}_{k})$
where $S$ is a scalar value and performing an operator commutation one gets  

\ba
S^{+}_{k}exp(-i\phi S^{z}_{k}) = exp(-i\phi(S^{z}_{k} + 1))S^{+}_{k} 
\label{eq3}   
\ea
and 

\ba
S^{-}_{k}exp(-i\phi S^{z}_{k}) = exp(-i\phi(S^{z}_{k} - 1))S^{-}_{k} 
\label{eq4}   
\ea
The $U_{k}(\phi)$ operator generates a translation of the Hamiltonian $H_{k-1,k}(\phi)$ to 
$H_{k,k+1}(\phi) = U^{\dagger}_{k}(\phi)H_{k-1,k}(\phi)U_{k}(\phi)$ and the wave functions are related 
by $|\Psi_{k,k+1}(\phi)\rangle = U^{\dagger}_{k}(\phi)|\Psi_{k-1,k}(\phi)\rangle$. The transformation 
induces a shift in the phase of the wave function which is given by the local order parameter 
defined by the BP, $\gamma_{k,l} = -i\int_{0}^{2\pi} d\phi \langle \Psi_{k,l}(\phi)|\partial_\phi|
\Psi_{k,l}(\phi)\rangle$. 

The corresponding Berry phases after and before the shift are then related by the expression 

\ba
\gamma_{k,k+1} = \gamma_{k-1,k} + \int \langle \Psi_{k-1,k}(\phi)|(S - S^{z}_{k})|
\Psi_{k-1,k}(\phi)\rangle
d \phi
\label{eq5}   
\ea
where $\gamma_{k-1,k}$ is the BP corresponding to $H_{k-1,k}$ and the second term on the r.h.s. is 
obtained from the phase factor\\ 

$\int \langle \Psi_{k-1,k}(\phi)|U_{k}(\phi)(d_{\phi}(U^{\dagger}_{k}(\phi))|
\Psi_{k-1,k}(\phi))\rangle d \phi$.\\


If the integral term $I_k $ in eq.(5) is such that $I_k =  2n\pi, n = integer$ and if the ground state
is non-degenerate before the translation then the BP is finite and the translational invariance of the 
system is realized. If the twist operation preserves the non-degeneracy of the ground state it means 
that there is a gap between this state and excited states. If $I_k \not= 2n\pi$ the translational 
invariance is violated. The BP induces a discontinuity which shows that the spectrum has no gap, 
hence it must show a gap closing.
 
In the following this test is applied to frustrated spin models in $d=2$ and $d=3$ dimensions in order 
to find out whether they can show or not a gap between the ground state and excited states.\\

{\it Application 1: the frustrated $2d$ system with an even and an odd number of legs}.

\begin{figure}
\epsfig{file=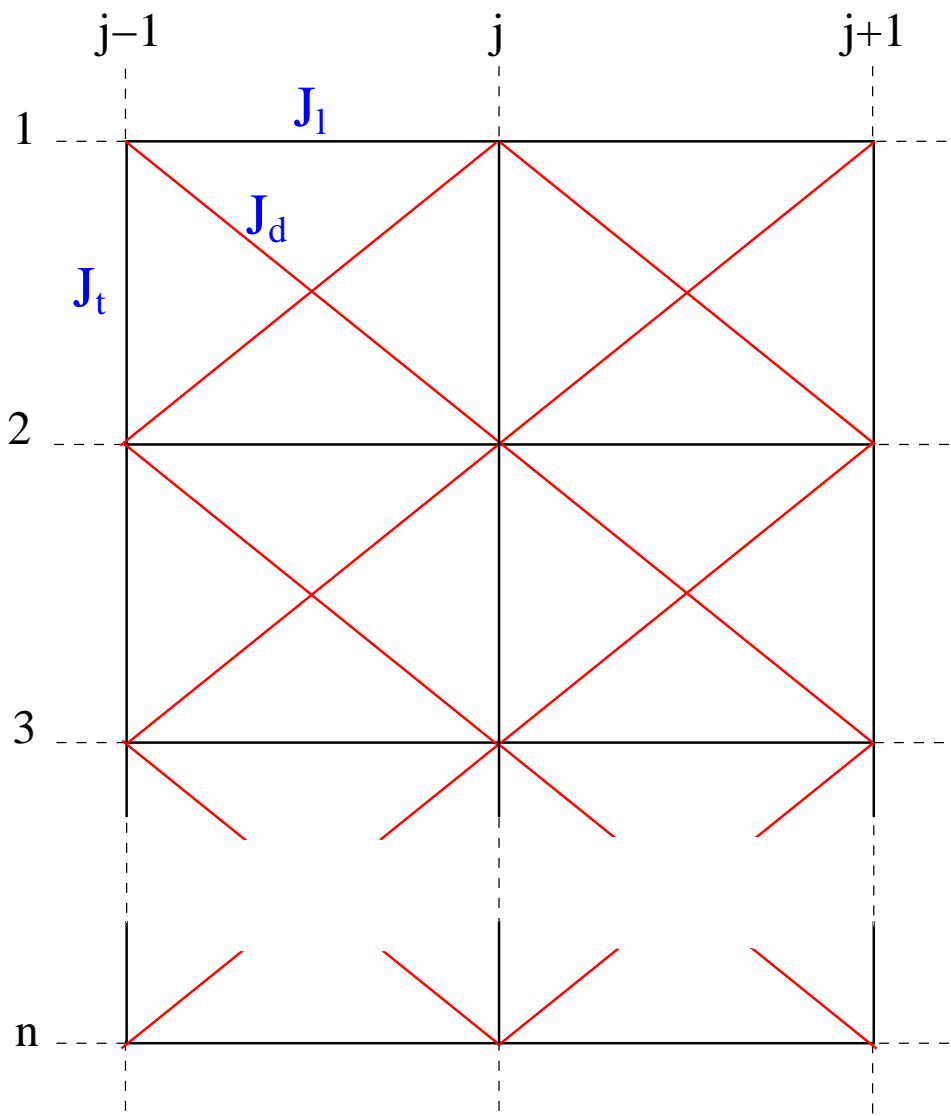,scale=1}
\caption{The 2d spin ladder with n legs. See the text.}
\label{fig1}
\end{figure}

Consider the frustrated ladder system with $n$ legs shown in fig.$1$ and the sites located at 
$[(j-1,i);(j,i);(j+1,i)]$ on the legs ($i = 1,...,n$).

The corresponding Hamiltonian reads 

\ba
H = H^{(l)} +H^{(t)} +H^{(d)} 
\label{6}
\ea
with 

\ba
H^{(l)} = \sum_{i=1}^{n}J_l(\vec S_{j-1,i}\vec S_{j,i} + \vec S_{j,i}\vec S_{j+1,i})
\label{7}
\ea
 
\ba
H^{(t)} = \sum_{i=1}^{n-1}J_t(\vec S_{j-1,i}\vec S_{j-1,i+1} + \vec S_{j,i}\vec S_{j,i+1} + 
\vec S_{j+1,i}\vec S_{j+1,i+1}) 
\label{8}
\ea

\ba
H^{(d)} = \sum_{i=1}^{n-1}J_d(\vec S_{j-1,i}\vec S_{j,i+1} + \vec S_{j-1,i+1}\vec S_{j,i} + 
\vec S_{j,i}\vec S_{j+1,i+1} + \vec S_{j,i+1}\vec S_{j+1,i})          ) 
\label{9}
\ea
where $J_l,J_t,J_d$ are coupling constants. Apply twists of angle $\phi_l$ and $\phi_d$
to the terms corresponding to sites ($j-1,j$) of $H^{(l)}$ and $H^{(d)}$. As done above shift
the twists from the ($j-1,j$) to the ($j,j+1$) sector by means of the unitary translation
operators $U_{j,i}(\phi_l)$ and $U_{j,i}(\phi_d)$, $[i=1,...,n]$. The wave functions corresponding 
to the sites ($j-1,i;j,i$) and ($j,i;j+1,i$), $(i = 1,...,n)$ are related by

\ba
|\Psi_{j,j+1}(\phi_l, \phi_d)\rangle = \prod_{i=1}^{n} U^{\dagger}_{j,i}(\phi_l)     
\prod_{i=1}^{n-1} U^{\dagger}_{j,i}(\phi_d)|\Psi_{j-1,j}(\phi_l, \phi_d)\rangle
\label{10}
\ea
where 

\ba
U_{j,i}(\phi_l)  = exp[i(S_{(j,i)}-S^{z}_{(j,i)})\phi_l]
\label{11}
\ea
and a similar expression for $U_{j,i}(\phi_d)$. $S_{(j,i)}$ is the length of a spin.

One mentions that the action of a translation operator on $H^{(t)}$ leaves this part of the 
Hamiltonian trivially invariant. Hence a twist cannot change the BP corresponding to this part of the
Hamiltonian when one moves from the ($j-1,j$) to the ($j,j+1$) sector. 

If one works out the Berry phase in the two sectors $S_{(j,i)}$ being half-integers the relation 
between the phases gets  

\ba\label{eq12}
\gamma^{(n)}(j,j+1) = \gamma^{(n)}(j-1,j) - 2\pi [\sum_{i=1}^{n} S_{(j,i)} + 2 \sum_{i=1}^{n-1} S_{(j,i)}]
\\ \nn
 - \int \sum_{k=l,d} \sum_{i_{k}}\langle \Psi_{j-1,j}(\phi_l,\phi_d)|S_{(j,i)}^{z}|\Psi_{j-1,j}
(\phi_l,\phi_d)
\rangle d\phi_k  
\ea
where the second and third term on the r.h.s. originate from the twist along the legs and the 
diagonals, the integral term contains the sum of both contributions and corresponds to the integral 
term in eq.(5).
 
If one works out the BP in the $(j-1,j)$ sector with $S_{(j,i)} = S$ this leads to 
\ba
\gamma^{(n)}(j,j+1) = \gamma^{(n)}(j-1,j) - 2\pi[n(S-<m>) + 2(n-1)(S-<m>)]
\label{13}
\ea
where the first term in the bracket corresponds to the contribution of the leg terms, the second to 
the contribution of the diagonal terms and $<m>$ is given by the integral term in Eq. (13). It is 
the magnetization per site for fixed $j$, the average being taken over the $j$s. In the following we
consider the case $S=1/2$.\\  


In a system with an even number $n$ of legs $<m>$ is an integer. $\Delta \gamma = \gamma^{(n)}(j-1,j) 
- \gamma^{(n)}((j,j+1) =  2\pi(3n-2)(S - <m>)$ is always a multiple of $2\pi$. In this case the 
translation conserves the gap.\\

In a system with an odd number of sites along the legs one must distinguish between two cases.

\begin{itemize}
\item $<m>$ is a half-integer and $\Delta \gamma$ can be a multiple of $2\pi$ if the number of legs 
$n$ is odd.
\item $<m>$ is an integer and $\Delta \gamma = \pi$ $mod(2\pi)$ if the number of legs $n$ is even.
\end{itemize}

Hence in the first case the spectrum may show a gap and no gap in the second case. 



{\it Application 2: frustrated $3d$ systems}.
 
The results obtained above can be extended to $3d$ systems in a straightforward way. We characterize 
the sites in the sectors $(j-1,j; j,j+1)$ by the indices $l$ and $m$ where $l$ stands for the legs 
in a given plane ($l = 1,..,n_l$), and $m$ ($m = 1,..,n_h$) for the planes. Define the Hamiltonian 
of the system with a maximal number of diagonal interactions: 
 
\begin{itemize}

\item $H^{(l)}(j-1,l,m;j,l,m)$ and the same with $(j,j+1)$: interactions along the legs in a plane
$(l)$ .

\item $H^{(t)}(j-1,l,m;j-1,l^{'},m)$ and the same with $(j)$ : interactions transverse to the legs 
in a plane $(l)$. They do not contribute to the BP.

\item $H^{(dl)}(j-1,l,m;j,l^{'},m)$ and the same with $(j \leftrightarrow j-1)$; $(j,j+1)$ and  
$(j \leftrightarrow j+1)$: interactions diagonal to the legs in a plane $(m)$.

\item $H^{(dt)}$: diagonal interactions between the planes along the transverse direction $(t)$. 
They do not contribute to the BP. 

\item $H^{(\bot)}$: interactions between the planes, along edges perpendicular to the planes. 
They do not contribute to the BP.   

\item $H^{(d \bot)}$: interactions $\bot$ to the planes, along the legs.

\item $H^{(lt)}$: interactions between the far vertices of a unit cubic cell. 

\end{itemize}

The spin-spin interactions work between nearest neighbours and may be characterized by different 
coupling strengths $(J_l,J_t,J_{dl},J_{dt},J_{\bot},J_{d \bot}, J_{lt})$.

One introduces twists on the different contributions of the total Hamiltonian $H_{tot}$ characterized
by different angles $\phi_l,...,\phi_{lt}$ and induces translations into the direction 
($(j,j-1)\rightarrow (j,j+1)$) by means of the unitary operators $U_{(j,k,l)}(\phi_l),...$. Following
the same lines as in the former $2d$ case one can evaluate the corresponding change of the BP 
from $\gamma_{3d}^{(-)}$ to $\gamma_{3d}^{(+)}$ when $(j,j-1)\rightarrow (j,j+1)$.\\

Taking account of all the phases accumulated through the different twists leads to a phase difference
$\Delta\gamma_{3d} = \gamma_{3d}^{(-)} - \gamma_{3d}^{(+)}$  

\ba\label{eq14}
\Delta\gamma_{3d} &=& 2\pi (S - <m>) [n_ln_h + 2n_l(n_h-1) + 2n_h(n_l-1)  
\\ \nn
& & + 4(n_l-1)(n_h-1)]  
\ea
where $<m>$ is the magnetization per site as defined through Eq. (12).   
In compact form

\ba
\Delta\gamma_{3d} = 2\pi (S - <m>) [9n_ln_h - 6(n_l + n_h) + 4] = 2\pi (S - <m>)C
\label{15}
\ea 

There are different cases to be considered when $S = 1/2$.

\begin{itemize}

\item If $<m>$ is a half-integer $\Delta\gamma_{3d}$ is a multiple of $2\pi$.

\item If $<m>$ is an integer $\Delta\gamma_{3d}$ is a multiple of $2\pi$ if $C$ is even
(one or both $n_l$ and $n_h$ are even).

\item If $<m>$ is an integer $\Delta\gamma_{3d}$ is a multiple of $\pi$ if $C$ is odd
(both $n_l$ and $n_h$ are odd).

\item If $<m>$ is neither integer nor half-integer $\Delta\gamma_{3d}$ takes an arbitrary 
value.

\end{itemize}

{\it Concluding remarks}.

The present work is an application of the Berry phase used as a topological test introduced 
by Y. Hatsugai~\cite{hat1}. The test is based on symmetry considerations and gives a $\it necessary$ 
condition for the existence or absence of a gap in the spectrum of frustrated spin systems.\\  

Once the spectrum of a system shows a gap before and after a local twist on the Hamiltonian one can 
conclude that the system is gapped.\\

This fact is consistent with former investigations and conjectures, among them those of 
refs.(2 - 12). It has been applied to $2d$ and $3d$ systems, but can in principle be extended to 
higher dimensions and different numbers of diagonal couplings.\\ 

Results are independent of the strengths of the coupling constants ${J_{kl}}$.\\ 

The theory fails to work if accidental level crossing points corresponding to possible phase transitions 
are present.\\ 


\newpage

\begin{thebibliography}{99}

\bibitem{hat1} Y. Hatsugai, J. Phys. Soc. Jpn. {\bf73} (2004) 2604, ibid.{\bf74} (2005)
1374 and {\bf75} (2006) 123601

\bibitem{hot} T. Hotta and E. Dagotto, Phys. Rev. Lett. {\bf 92} (2004) 227201

\bibitem{tas}L. Tassini, F. Venturi, Q.-M. Zhang, R. Hackl, N. Kikugawa and T. Fujita,
cond-mat/0406169

\bibitem{hay} S. M. Hayden, H. A. Mook, Pengcheng Dai, T. G. Perring and F. Dogan, 
 Nature {\bf 429} (2004) 531

\bibitem{tran} J.M. Tranquada, H. Woo, T.G. Perring, H. Goka, G. D. Gu, G. Xu, M. Fujita and 
K. Yamada, Nature {\bf 429} (2004) 534 

\bibitem{dag} E. Dagotto and T. M. Rice, Science {\bf 271} (1996) 618

\bibitem{dell} S. Dell'Aringa, E. Ercolessi, G. Morandi, P. Pieri, M. Roncaglia, Phys.Rev. 
Lett. {\bf 78} (1997) 2457 

\bibitem{azu} M. Azuma, Z. Hiroi, M. Takano, K. Ishida and Y. Kitaoka, Phys.Rev. Lett.
{\bf 73} (1994) 3463 

\bibitem{hald1} F. D. M. Haldane, Phys. Rev. Lett. {\bf 50} (1983) 1153

\bibitem{hald2} F. D. M. Haldane, Phys. Lett. {\bf 93A} (1983) 464

\bibitem{ric1} J. Richert, arXiv: cond-mat/0510343

\bibitem{ric2} J. Richert, O. G\"uhne, to appear in Physica Status Solidi {\bf B}

\bibitem{wen1} X. G. Wen, Phys. Rev. {\bf B40} (1989) 7387

\bibitem{wen2} Y. Ran, X. G. Wen, Phys. Rev. Lett. {\bf 96} (2006) 026802

\bibitem{wen3} M. Levi, X. G. Wen, Phys. Rev. Lett. {\bf 96} (2006) 110405

\bibitem{kit1} A. Kitaev, J. Preskill, Phys. Rev. Lett. {\bf 96} (2006) 110406

\bibitem{berr} M. V. Berry, Proc. R. Soc. Lond. {\bf A 392} (1984) 45

\bibitem{goss} B. Levi Goss, Phys. Today  {\bf 46} (1993) 17

\bibitem{wilc} F. Wilczek, A. Zee, Phys. Rev. Lett. {\bf 52} (1984) 2111

\bibitem{zhan} Y. Zhang, Y. W. Tan, H. L. Stormer, P. Kim, Nature{\bf 438} 201

\bibitem{zhao} Xuean Zhao, T. C. Au Teung, Ya-Bin Yu, C. H. Kam, You-Quan Li,
Eur. Phys. Lett. {\bf 82} (2008) 1005

\bibitem{ryu} S. Ryu, Y. Hatsugai, Phys. Rev. {\bf B73} (2006)245115 

\bibitem{binz} Bin Zhou, Chao-Xing Liu, Shun-Qing Shen, Europhys. Lett. {\bf 79} (2007)
47010 and arXiv:cond-mat/0705.3728 

\bibitem{shil} Shi-Liang Zhu, arXiv:quant-ph/0803.1914

\bibitem{hat2} Yasuhiro Hatsugai, arXiv:cond-mat/0607024 and conference contribution (HFM 2006)

\bibitem{hat3} Y. Hatsugai, J. Phys. Condens. Matter {\bf 19} (2007) 145209

\bibitem{hat4} T. Hirano, H. Katsura, Y. Hatsugai, Phys. Rev. {\bf B77} (2008) 094431 

\bibitem{hat5} Yasuhiro Hatsugai, arXiv:cond-mat/0603230 and J. Phys. Soc. Jpn. {\bf75} (2006) 123601 

\bibitem{hat6} T. Hirano, H. Katsura, Y. Hatsugai, arXiv:cond-mat/0803.3185  

\end{thebibliography}
\end{document}